\documentclass{aastex}
\usepackage{lscape,afterpage}

\usepackage{graphicx}

\usepackage{spr-astr-addons}
\usepackage{experimental}

\shorttitle{Ara~OB1 - interstellar medium}
\shortauthors{Henderson et al.}

\begin{document}

\title{The interstellar medium towards the Ara~OB1 region.}

\author{
Christopher D.\ Henderson\altaffilmark{1}, Barry Y.\ Welsh\altaffilmark{2} and John B.\
Hearnshaw\altaffilmark{1} }

\altaffiltext{1}{Department Physics $\&$ Astronomy, University of Canterbury, Christchurch,
New Zealand; cdh29@student.canterbury.ac.nz} \altaffiltext{2}{Experimental Astrophysics
Group, Space Sciences Laboratory, University of California, 7 Gauss Way,  Berkeley, CA
94720, USA}

\begin{abstract}
We present high resolution ($R \sim$ 4\,km\,s$^{-1}$) absorption measurements of the
interstellar NaI and CaII lines measured towards 14 early-type stars of distance
123\,pc--1650\,pc, located in the direction of the Ara~OB1 stellar cluster. The line
profiles can broadly be split into four distinct groupings of absorption component
velocity, and we have attempted to identify an origin and distance to each of these
interstellar features. For gas with absorption covering the velocity range
$-10$\,km\,s$^{-1}$ $<$ $V_{\mathrm{helio}}< +10$\,km\,s$^{-1}$, we can identify the
absorbing medium with local gas belonging to the Lupus-Norma interstellar cavity located
between 100 and 485\,pc in this galactic direction. Gas with velocities spanning the range
$-20$\,km\,s$^{-1}< V_{\mathrm{helio}} < +20$\,km\,s$^{-1}$ is detected towards stars with
distances of 570--800\,pc. We identify a wide-spread interstellar feature at
$V_{\mathrm{helio}} \sim -15$\,km\,s$^{-1}$ with the expanding HI shell called
GSH~337+00$-$05, which is now placed at a distance of $\sim$ 530\,pc.

Gas with velocities in the range  $-45$\,km\,s$^{-1} < V_{\mathrm{helio}} <
 -25$\,km\,s$^{-1}$ is only detected towards stars with distances $>$ 1\,kpc, which places
this absorbing gas at a similar distance to that of the Ara~OB1 stellar cluster. The
similarity between the radial velocity of the stellar cluster and that of the absorbing gas
suggests that the cluster stars may be presently embedded within, and traveling through,
this interstellar cloud complex. Absorption at velocities $<$ $-50$\,km\,s$^{-1}$ is seen
towards only two stars, HD 154873 and HD 150958, both of which have distances $>$ 1300\,pc.
Although the NaI/CaII for this component is consistent with the presence of interstellar
shocks, we believe that it is unlikely that this highly negative velocity gas is physically
associated with the Ara~OB1 cluster itself, and its origin may be linked to the more
distant (and larger) Sco~OB1 cluster.
\end{abstract}


\section{Introduction}
The Ara~OB1 association is located in the Carina-Sagittarius spiral arm  ($\it l \sim 337
^{\circ}$, $\it b \sim -1.5^{\circ}$) and contains several clusters of  hot young stars
that include NGC 6167, 6193 and 6204, as well as the molecular cloud/HII region complexes
of RCW 108 and 107. With a distance of $\sim$ 1320\,pc \citep{herbst77} this complex lies
slightly foreground and adjacent to the larger Sco OB1 association. The interstellar
sight-line to this region of the Galaxy is known to be complex, and it contains several gas
and dust clouds with velocities covering the $-50$ to $+10$\,km\,s$^{-1}$ range
\citep{arnal03}. A constant reddening of $\sim$ 0.2 mag has been found in this direction
reaching to a distance of $\sim$ 1\,kpc \citep{herbst77}, with the majority of this
interstellar dust residing within 200\,pc of the Sun. Beyond this local extinction there
appears to be a clear interstellar region, $\sim$ 1\,kpc wide, which is the well-known
`inter-arm' gap that separates the local spiral arm from the adjacent Carina-Sagittarius
arm.

The young stars in Ara~OB1 are known to have a profound effect on the surrounding
interstellar medium (ISM), with their out-flowing stellar winds imparting pressure to form
expanding shells of gas. Many such interstellar features have been discovered from HI and
H$\alpha$ observations of the area. For example, \cite{arnal87} have found an expanding HI
emitting shell centered on NGC 6167 with a kinematic distance of 1400$\pm 200$\,pc  and an
outer radius of $\sim 50$\,pc. Subsequent HI observations of this region by \cite{rizzo94}
have further revealed a pervasive two-shell structure with gas cloud LSR velocities
centered at  $-16$\,km\,s$^{-1}$ and $-34$\,km\,s$^{-1}$, each with respective kinematic
distances of 1.4 and 2.5\,kpc. The nearer gas shell was speculated to be formed as a result
of stellar winds (and/or supernova explosions), which sequentially triggered the star
formation process in Ara~OB1 after running into a nearby molecular cloud less than $2
\times 10^{6}$ years ago \citep{wald99}. H$\alpha$  observations of this region by
\cite{georg96} have also revealed similarly pervasive emission features with LSR velocities
of $-28$\,km\,s$^{-1}$ and $-39$\,km\,s$^{-1}$.

More recent HI observations for the Southern Galactic Plane Survey by \cite{mcclure02} have
revealed a large 10$^{\circ}$ angular diameter shell, named GSH~337+00$-$05, surrounding
this entire region that is centered approximately on the galactic plane. The authors derive
a systemic LSR velocity of $-5$\,km\,s$^{-1}$, an expansion velocity of 9\,km\,s$^{-1}$
and an energy of $\sim 1.6 \times 10^{44}$\,J for this large interstellar feature. They
place the distance of this gas shell at 570$\pm 900$\,pc, which would imply that it lies
considerably in front of the Ara~OB1 cluster, although the expansion energy of the shell is
consistent with its formation by a young association of hot stars (such as those of
Ara~OB1). Interstellar NaI and CaII absorption observations of the sight-lines towards
early-type stars in both Sco OB1 ($\it l = 343^{\circ}$, $\it b = +1^{\circ}$) by
\cite{crawford89} and in Ara~OB1 by \cite{white85} have revealed several gas cloud
components which show several well-defined groupings with LSR velocities in the $-15$ to
$-50$\,km\,s$^{-1}$ range, similar to that found from the HI data. However, these
negative velocity absorption components have been associated with interstellar gas at a
distance $>$ 1\,kpc that has supposedly been disrupted by the stellar winds of the two
aforementioned OB associations. Thus, the interpretations of the radio HI emission and the
ground based absorption data are presently in disagreement.

In this paper we present new high resolution ($R = 70\,000$ or velocity resolution $\sim
4$\,km\,s$^{-1}$) NaI and CaII absorption measurements towards 14 early-type stars located
in the galactic direction towards Ara~OB1. From these new, and previous, interstellar
observations we can place a most probable distance of $\sim 530$\,pc to the large
GSH~337+00$-$05 HI shell, in contradiction to its physical association with the Ara~OB1
cluster as argued previously by several authors. Instead, we argue that gas with absorption
velocities in the range $-45$\,km\,s$^{-1} < V_{\mathrm{helio}} < -25$\,km\,s$^{-1}$ is
formed at a distance of $\sim  1$\,kpc, and is far more likely to be linked to the
interaction of the Ara~OB1 stellar cluster with the ambient ISM.

\section{Observations and data analysis}

Interstellar NaI D-line doublet (5890\,\AA) and CaII (3933\,\AA) absorption line
measurements were obtained for the 14 early-type stars given in Table 1, which all lie
within a radius of $\sim  7^{\circ}$ of the central position of the GSH~337+00$-$05 shell.
We also list the galactic position, visual magnitude, spectral type, reddening (colour
excess) and distances of these stars. The list (presented in increasing stellar distance)
includes targets lying both in front of and beyond the nominal 1320\,pc distance of the
Ara~OB1 association. In Figure 1 we show the position of all these (numbered) targets with
respect to an HI image of the GSH~337+00$-$05 shell \citep{mcclure02}. The majority of our
targets lie within $\pm 2^{\circ}$ of the galactic plane and most have sight-lines that are
coincident with the brightest emission features of the $-5.5$\,km\,s$^{-1}$ HI gas shell.

Well determined distances for the stars are an important parameter for this study, but
unfortunately only one target (HD 147971) has an accurate Hipparcos trigonometric parallax
measurement. For the remainder of the targets we present spectro-photometric distance
estimates that have been derived using their spectral type, absolute magnitude and colour
excess, as reported in the references listed  in Table 1 for each star. The typical error
for the majority of these derived distances is $\sim 25$ per cent \citep{hunter06}.

 \begin{figure*}[htb]
\includegraphics[width=16cm]{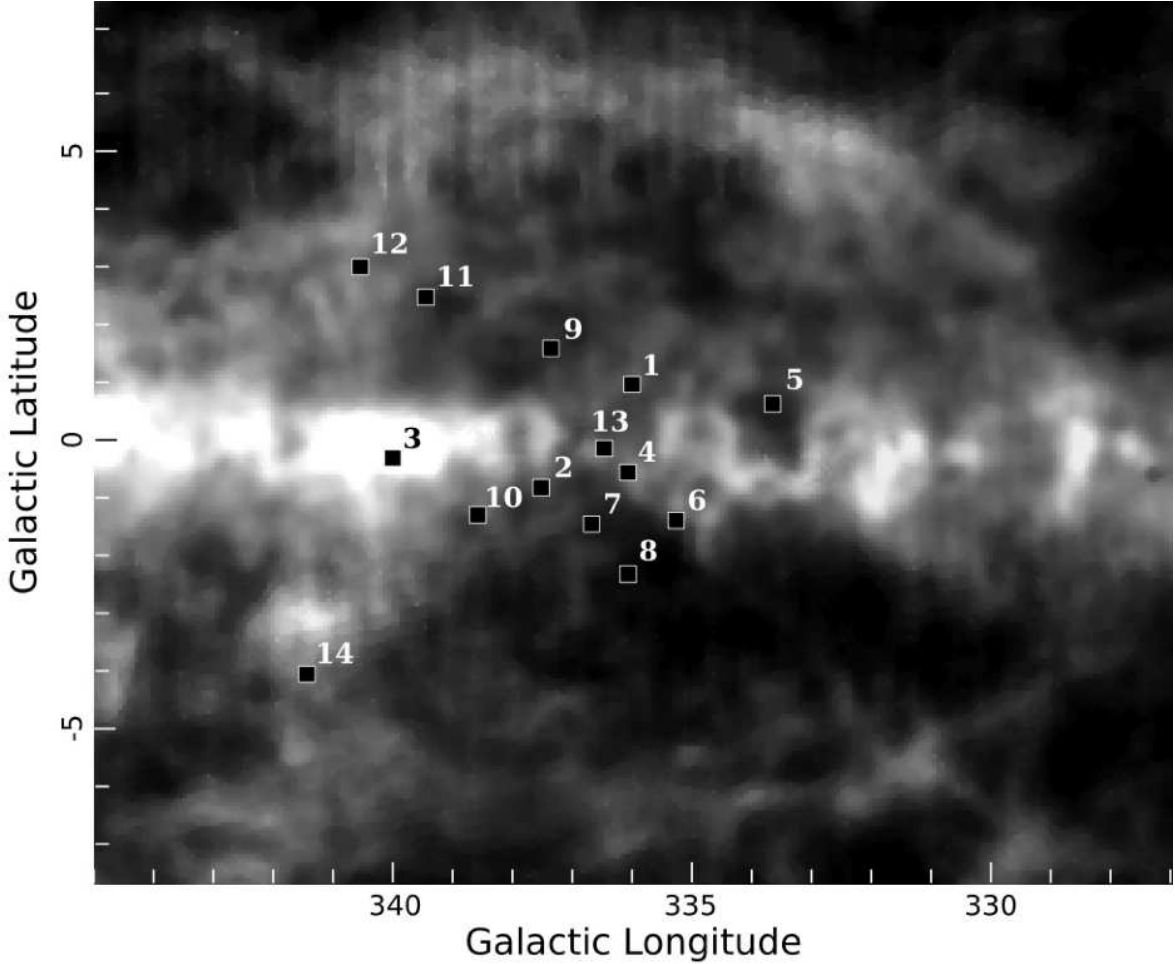}
\caption{The GSH~337+00$-$05 HI shell from the Southern Galactic Plane Survey
\citep{mcclure02}. Superimposed are the 14 target stars used for absorption measurements.
Numbering of the targets follows the list given in Table 1.} \label{Figure 1}
\end{figure*}

The spectroscopic observations were carried out during April 2006, June 2006 and July 2007
using the 1.0-m McLellan telescope at the Mt John Observatory in New Zealand. The HERCULES
vacuum fibre-fed \'echelle spectrograph \citep{hearnshaw02} was used with the 100 micron
fibre and 50 micron slit to give a resolving power of $R = 70\,000$ ($\sim
4$\,km\,s$^{-1}$). The data were recorded with a Fairchild 4k $\times$ 4k CCD in a Spectral
Instruments camera, and wavelength calibrations were obtained from thorium-argon
hollow-cathode lamp spectra taken at the beginning and end of each night. The two spectral
orders of interest were extracted from the raw CCD images using software routines written
by J.\ Skuljan (University of Canterbury, Christchurch). Standard \'echelle reduction
procedures were employed that included background and cosmic-ray subtraction, spectral
order tracing and extraction, flat-fielding and wavelength calibration. For the case of the
NaI D-line doublet order, each stellar spectrum was divided by that of the bright and
nearby star $\alpha$ Eri in order to remove the narrow telluric water vapor lines that
contaminate this wavelength region. The majority of the stellar spectra were well-exposed
with typical S/N ratios in excess of 25:1 (less for the CaII data), and all velocities
subsequently reported in this paper are in the heliocentric frame of reference (for
comparison, $V_{\mathrm LSR} = V_{\mathrm{helio}} + 3$\,km\,s$^{-1}$).

The interstellar absorption profiles of the two NaI D-lines and the CaII K-line  were fit
using  a dedicated software package described in \cite{sfeir99}. This fit process entailed
determining the local stellar continua for each absorption line using a multi-order
polynomial fitting procedure. The resultant residual intensity profiles, whose equivalent
widths are listed in Table 1 (with typical measurement errors of $\pm 10$ per cent), were
then fit with multiple gas cloud components, each with a theoretical best-fit of cloud
component velocity ($V$), cloud doppler velocity dispersion ($\it b$) and cloud component
column density ($N$). The CaII absorption profiles were less saturated than the NaI lines
and were fit prior to the NaI lines for each sight-line. The CaII model fit was then used
as a basis for the subsequent fitting of the NaI D-lines. Each model fit procedure used the
$\it minimum$ number of cloud components that were statistically significant using the
criterion given in \cite{vallerga93}. The two NaI D-lines were fit simultaneously in order to better constrain the model fits, whereas each CaII K-line was fit independently. The
resultant absorption model fit parameters for each star are listed in Table 2, and the
individual fits superimposed on the data points are shown in Figures 2--4. We have also
limited the maximum gas cloud temperatures (i.e. the derived $\it b$-values) in the Na D-line fits
to $15\,000$\,K and $20\,000$\,K for the CaII lines, in order to comply with the
expected physical conditions in the diffuse ISM.

The low S/N ratio in the spectrum of HD 149019 at the CaII K-line means that the fit is probably over-interpreted. In that case knowledge of possible velocity components was gleaned from other distant stars and used to fit the line with the minimum number of components.

The line-fit procedure works well for unsaturated absorption components, but  large errors
can occur in the derived column densities for the central cores of  highly saturated
absorption lines. Unfortunately, since the majority of our targets lie along distant and
high column density sight-lines, significant line saturation is present for cloud
components with velocities in the $-20$ to $+10$\,km\,s$^{-1}$ range. However, when
components reside in the (red or blue) wings of the absorption profiles their column
densities are better constrained. Thus, for most of the absorption profiles, only the very
central components cannot be fit with certainty and these components are identified as such
in Table 2.  Much of the following analysis of our observations fortunately relies on the
velocities, and not the column densities, of the numerous absorption components.

\begin{deluxetable}{lccccccccc}
\tabletypesize{\scriptsize} \rotate \tablecaption{Stellar target information}
\tablewidth{0pt} \tablehead{ \colhead{Star} & \colhead{$\it(l,b)$} & \colhead{$m_{\rm v}$}
& \colhead{Sp} & \colhead{$E(B-V$)} & \colhead{Distance} & \colhead{Reference} &
\colhead{EW (D2)} & \colhead{EW (D1)} & \colhead{EW (CaII)}
\\
&&&&&\colhead{(pc)} & & \colhead{m\AA\ } & \colhead{m\AA\ } & \colhead{m\AA\ }
}
\tablecolumns{10}
\startdata
(1) HD 147971&$ (336.0,+1.0)$&4.5&B4V&0.09&123$\pm15$ & (1)&115&100&18\\
(2) HD 150083&$ (337.5,-0.9)$&7.3&B5III&0.20&485$\pm125$ & (2) &240&205&80\\
(3) HD 151113&$ (339.9,-0.2)$&6.7&B5II/III&0.17&570$\pm145$ & (3)&405&335&145\\
(4) HD 148974&$ (336.1,-0.5)$&6.9&A7III&0.05&600$\pm350$& (1)&235&185&145\\
(5) HD 146444&$ (333.7,+0.7)$&7.6&B2V&0.25&800$\pm200$& (3)&285&235&105\\
(6) HD 149019&$ (335.2,-1.4)$&7.5&A0I&0.84&1108$\pm110$& (4)&605&560&235\\
(7) HD 150041&$ (336.7,-1.5)$&7.1&B1/B2I&0.39&1155$\pm115$& (4)&470&410&180\\
(8) HD 150136&$ (336.1,-2.2)$&5.7&O3&0.46&1155$\pm115$&(4)&595&500&235\\
(9) HD 148379&$ (337.3,+1.6)$&5.4&B1.5I&0.61&1180$\pm295$&(5)&745&680&-\\
(10) HD 150958&$ (338.6,-1.2)$&7.3&O6e&0.61&1300$\pm325$&(4)&880&745&-\\
(11) HD 149038&$ (339.4,+2.5)$&4.9&B0I&0.31&1380$\pm345$&(6)&320&300&-\\
(12) HD 149404&$ (340.5,+3.0)$&5.5&O9I&0.66&1380$\pm345$&(6)&610&580&315\\
(13) HD 148937&$ (336.4,-0.2)$&6.8&O6.5&0.65&1380$\pm345$&(5)&700&650&-\\
(14) HD 154873&$ (341.4,-4.1)$&6.7&B1Ib&0.47&1650$\pm415$&(5)&655&615&265\\
\enddata
\tablerefs{
(1) Perryman 1997;
(2) Kaltcheva $\&$ Georgiev 1992;
(3) Kozoc 1985;
(4) Kharchenko et al. 2005;
(5) Hunter et al. 2006;
(6) Crawford, Barlow and Blades 1989;
}
\end{deluxetable}

 \begin{figure*}
\center
\includegraphics[width=16cm]{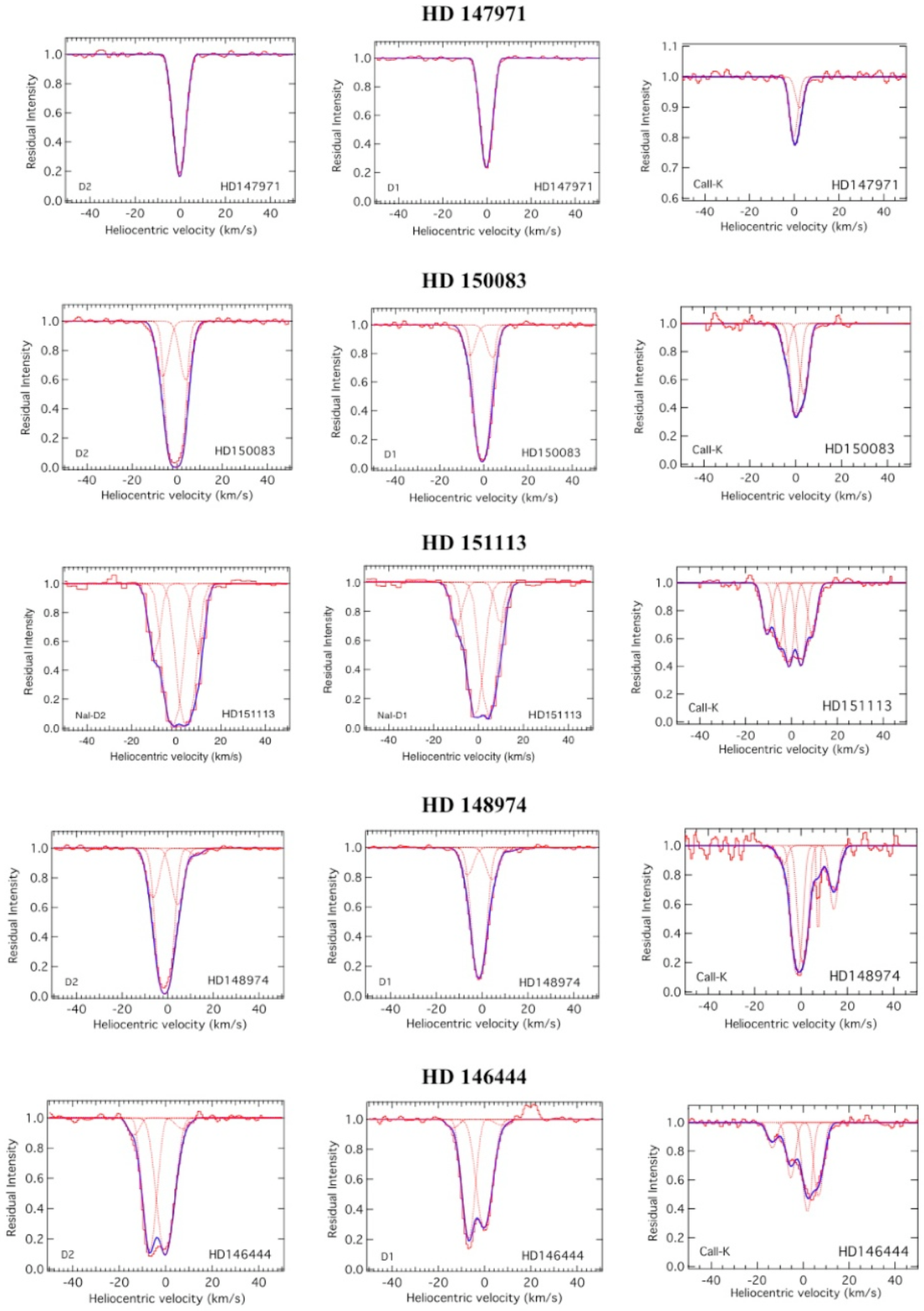}
\caption{NaI (D2 and D1) and CaII-K absorption line profiles for targets $\#$ 1 -- 5.
Dotted lines represent the individual components required to fit the entire line absorption
profile (bold line).} \label{Figure 2}
\end{figure*}

 \begin{figure*}
\center
\includegraphics[width=16cm]{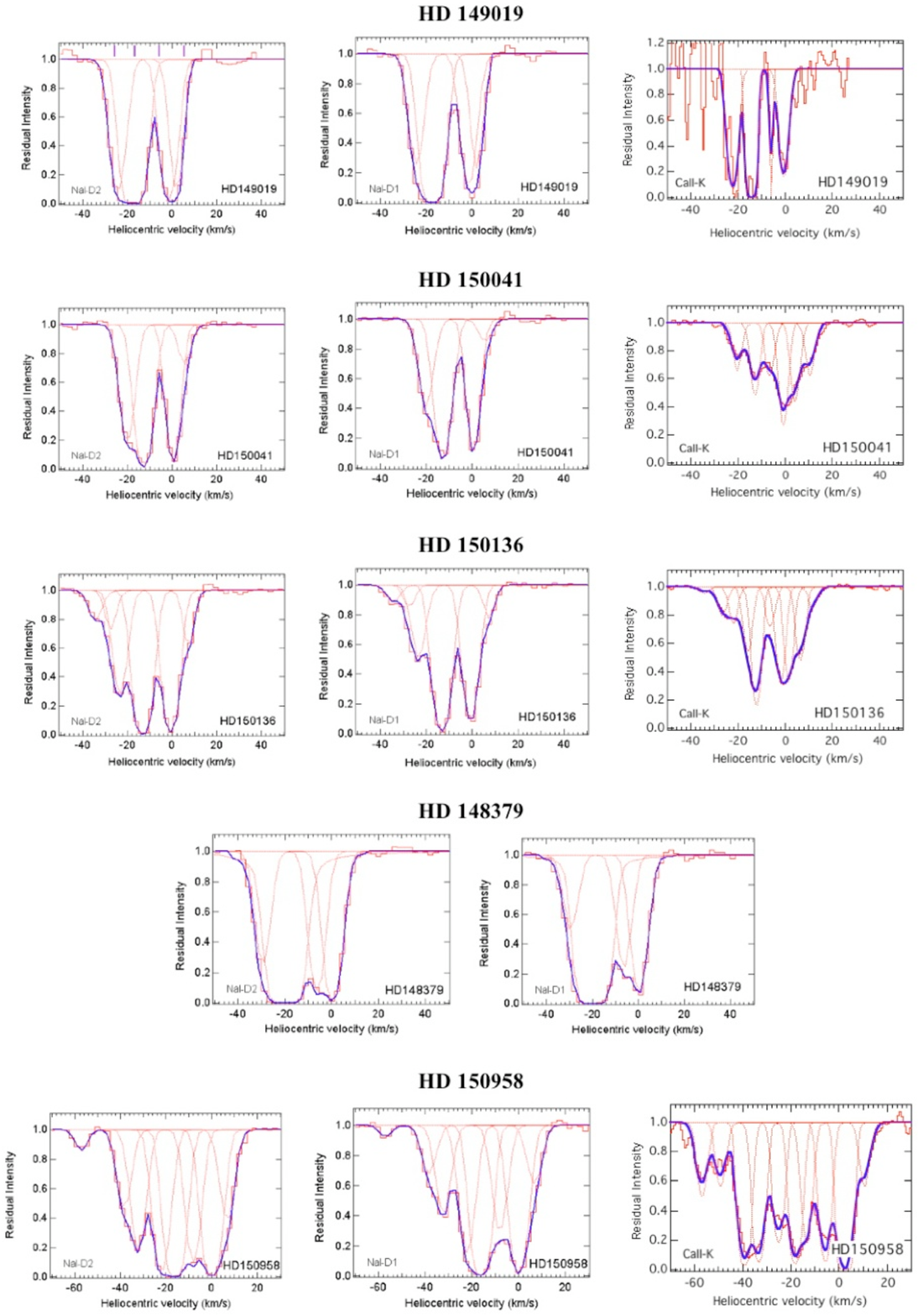}
\caption{NaI (D2 and D1) and CaII-K absorption line profiles for targets $\#$ 6 -- 10.
Dotted lines represent the individual components required to fit the entire line absorption
profile (bold line).} \label{Figure 3}
\end{figure*}

 \begin{figure*}
\center
\includegraphics[width=16cm]{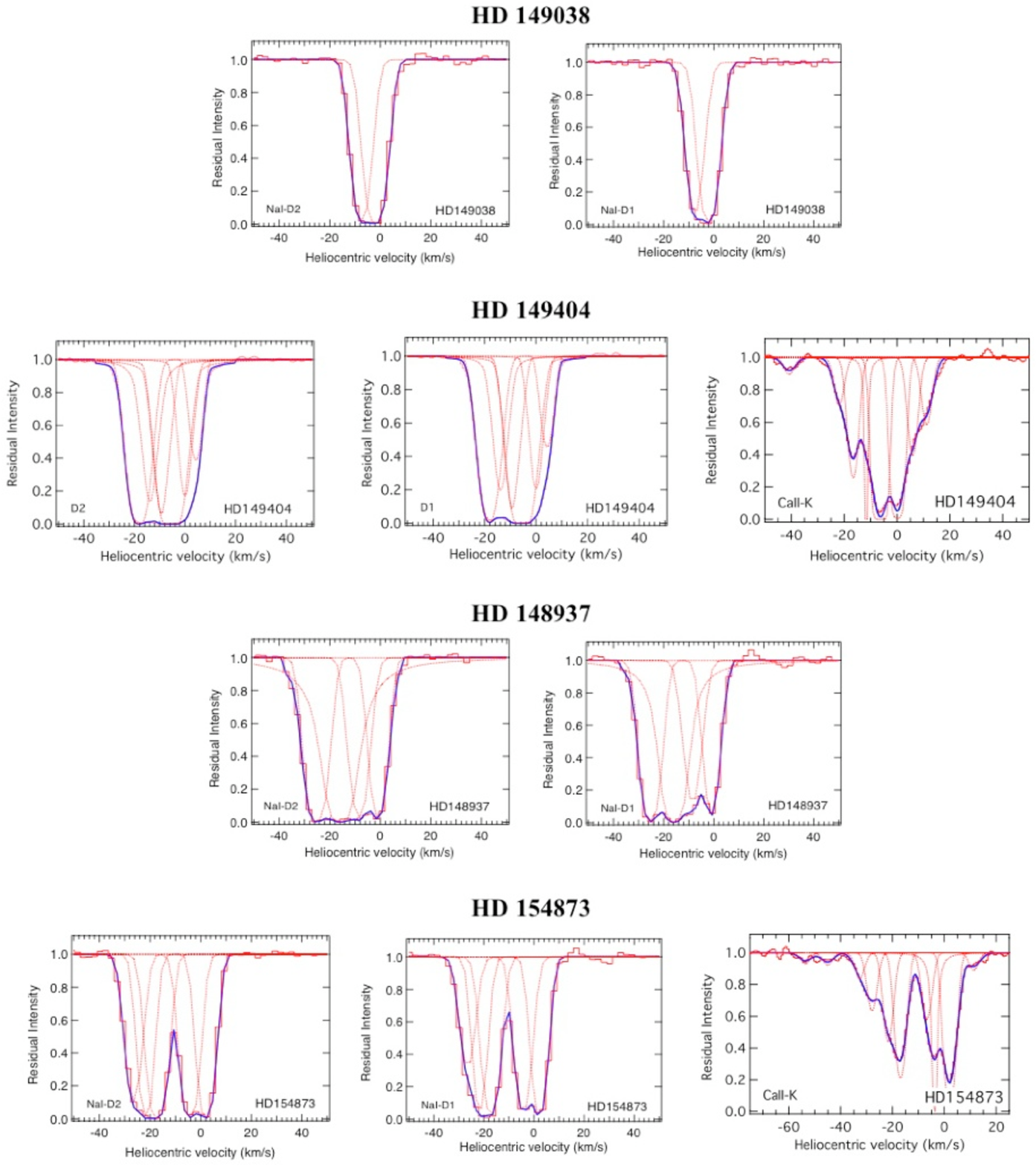}
\caption{NaI (D2 and D1) and CaII-K absorption line profiles for targets $\#$11 -- 14.
Dotted lines represent the individual components required to fit the entire line absorption
profile (bold line).} \label{Figure 4}
\end{figure*}

{
 \renewcommand{\arraystretch}{.7}
\begin{table*}[htbp]
\caption{\label{obslog}NaI and CaII absorption line best-fit parameters}
\begin{flushleft}
\tiny
\begin{tabular}{lcccccccccc}
\hline
\hline
&&&&&&&&&& \\
I.D. $\#$&Star& $V_{\mathrm{helio}}$ &$\it b$&$N$&&& $V_{\mathrm{helio}}$ &$\it b$&$N$& NaI/CaII \\
&&km s$^{-1}$&&(10$^{10}$ cm$^{-2}$)&&&km s$^{-1}$&&(10$^{10}$ cm$^{-2}$)& \\
\hline &&&&&&&&&& \\
1&{\bf HD 147971} &&&& \\
&...NaI...&&&&&...CaII...&+2.5&2.5&7$\pm1$&-\\
&&-0.3&2.9&130$\pm15$&&&-0.2&2.4&15$\pm2$&8.7 \\
2&{\bf HD 150083} &&&& \\
&...NaI...&+3.9&3.3&33$\pm$5&&...CaII...&+3.5&2.5&47$\pm8$&0.7 \\
&&-0.7&3.3&390$\pm140$&&&-0.2&2.4&70$\pm10$&5.6 \\
&&-6.5&3.3&30$\pm6$&&&-4.4&2.5&17$\pm3$&1.8  \\
3&{\bf HD 151113} &&&& \\
&...NaI...&+10.1&2.0&44$\pm$10&&...CaII...&+9.5&2.4&33$\pm6$&1.3 \\
&&+4.6&3.3&445$\pm150$&&&+4.2&2.5&60$\pm10$&7.4 \\
&&-1.6&3.3&455$\pm150$&&&-1.0&2.5&65.0$\pm10$&7.0  \\
&&&&&&&-5.6&2.5&40$\pm10$&- \\
&&-9.7&2.2&45$\pm10$&&&-10.8&2.5&32$\pm5$&1.4  \\
4&{\bf HD 148974} &&&& \\
&...NaI...&+12.5&3.3&3$\pm$1&&...CaII...&+14.1&2.3&35$\pm5$&0.09 \\
&&+4.3&3.3&30$\pm8$&&&+7.0&0.6&19$\pm8$&1.6 \\
&&&&&&&+1.0&2.5&110$\pm40$&- \\
&&-1.4&3.3&265$\pm80$&&&-1.8&2.5&140$\pm40$&1.9 \\
&&-6.5&3.3&25$\pm5$&&&-7.9&2.5&10$\pm2$&2.5 \\
5&{\bf HD 146444} &&&& \\
&...NaI...&+7.3&2.5&6$\pm$1&&...CaII...&+6.8&2.5&48$\pm8$&0.1 \\
&&-0.1&3.3&210$\pm50$&&&+1.5&2.5&65$\pm15$&3.2 \\
&&-6.9&1.8&370$\pm80$&&&-5.5&2.5&35$\pm6$&10.6 \\
&&-13.8&1.3&8$\pm1$&&&-13.3&2.4&13$\pm2$&0.6 \\
6&{\bf HD 149019} &&&& \\
&...NaI...&+1.9&2.4&220$\pm$40&&...CaII...&-0.5&2.5&120$\pm10$&1.8 \\
&&-1.6&3.3&370$\pm80$&&&-5.0&0.5&60$\pm15$&6.2 \\
&&-17.0&3.1&$>$ 1000**&&&-15.0&1.7&860$\pm150$&n/a \\
&&-24.0&3.3&210$\pm40$&&&-22.0&2.5&175$\pm40$&1.2\\
7&{\bf HD 150041} &&&& \\
&...NaI...&&&&&...CaII...&+10.5&2.5&32$\pm4$&- \\
&&+5.5&2.8&22$\pm4$&&&+4.3&2.5&55$\pm15$&0.4 \\
&&&&&&&-6.5&2.5&32$\pm6$&- \\
&&-12.5&3.3&550$\pm100$&&&-12.8&2.5&48$\pm10$&1.1\\
&&-20.0&3.3&140$\pm20$&&&-20.3&2.5&27$\pm4$&5.2\\
8&{\bf HD 150136} &&&& \\
&...NaI...&&&&&...CaII...&+11.7&2.5&10$\pm2$&- \\
&&+7.3&2.1&30$\pm5$&&&+6.5&2.5&49$\pm10$&0.6 \\
&&-0.5&3.1&465$\pm150$&&&+1.5&2.5&70$\pm25$&6.6 \\
&&&&&&&-2.0&2.5&78$\pm25$&-\\
&&&&&&&-6.5&2.5&25$\pm5$&-\\
&&-13.0&3.3&980$\pm400$**&&&-12.2&2.5&120$\pm50$&8.2**\\
&&&&&&&-16.0&2.5&40$\pm15$&-\\
&&-23.5&3.3&100$\pm35$&&&-22.0&2.5&20$\pm5$&5.0\\
&&-27.5&2.4&20$\pm8$&&&-26.6&2.4&16$\pm4$&1.3\\
&&&&&&&-34.0&2.4&3$\pm1$&-\\
9&{\bf HD 148379} &&&& \\
&...NaI...&0.3&3.3&460$\pm80$&&...CaII...&&&&- \\
&&-6.3&3.3&225$\pm40$&&&&&&- \\
&&-20.0&3.3&$>$ 1000**&&&&&&- \\
&&-29.5&3.3&105$\pm15$&&&&&&-\\
10&{\bf HD 150958} &&&& \\
&...NaI...&+7.0&3.3&75$\pm10$&&...CaII...&+10.0&2.6&50$\pm5$&1.5 \\
&&+1.0&3.3&905$\pm180$&&&+2.5&2.5&705$\pm120$&1.3\\
&&-8.0&3.2&195$\pm40$&&&-6.5&2.5&210$\pm40$&0.9\\
&&&&&&&-12.5&2.6&125$\pm30$&-\\
&&-16.0&2.7&$>$1000**&&&-18.0&2.5&230$\pm80$&-\\
&&-23.0&3.3&315$\pm65$&&&-24.5&2.5&125$\pm30$&2.5\\
&&-32.0&3.3&140$\pm35$&&&-33.0&2.5&220$\pm60$&0.6\\
&&-38.5&3.3&58$\pm15$&&&-39.0&2.5&265$\pm60$&0.2\\
&&&&&&&-49.0&2.5&42$\pm5$&-\\
&&-57.1&2.7&10$\pm2$&&&-57.1&2.5&50$\pm5$&0.2\\
11&{\bf HD 149038} &&&& \\
&...NaI...&-1.9&3.2&990$\pm200$&&...CaII...&&&&- \\
&&-7.5&3.0&505$\pm100$&&&&&&- \\
12&{\bf HD 149404} &&&& \\
&...NaI...&&&&&...CaII...&+11.2&2.5&35$\pm4$&- \\
&&+4.5&1.1&255$\pm45$&&&+5.9&2.5&60$\pm8$&4.3 \\
&&+0.0&1.1&$>$1000**&&&+0.2&2.5&345$\pm100$&- \\
&&-5.4&2.6&$>$1000**&&&-6.4&2.5&680$\pm200$&- \\
&&-9.5&1.3&$>$ 1000**&&&-11.5&2.5&$>$1000**&-\\
&&-14.5&0.8&$>$ 1000**&&&-16.0&2.5&90$\pm20$&- \\
&&-18.5&3.2&$>$1000***&&&-21.5&2.5&25$\pm3$&- \\
&&&&&&&-40.8&2.5&7.5$\pm1$&- \\
13&{\bf HD 148937} &&&& \\
&...NaI...&-0.9&2.5&705$\pm80$&&...CaII...&&&&- \\
&&-8.5&3.3&355$\pm65$&&&&&&- \\
&&-16.0&1.8&$>$1000**&&&&&&- \\
&&-25.5&3.0&$>$1000**&&&&&&- \\
14&{\bf HD 154873} &&&& \\
&...NaI...&&&&&...CaII...&+11.5&2.5&8$\pm2$&- \\
&&+2.2&2.7&695$\pm75$&&&+2.1&2.5&180$\pm20$&3.9 \\
&&-4.5&3.1&470$\pm55$&&&-4.0&0.5&$>$1000**&$<$0.5** \\
&&&&&&&-7.0&2.5&36$\pm5$&- \\
&&-17.0&3.3&575$\pm65$&&&-17.0&2.5&105$\pm15$&5.5 \\
&&-21.5&2.4&510$\pm65$&&&-21.5&2.5&55$\pm10$&9.3 \\
&&-26.5&3.3&165$\pm30$&&&-28.0&2.5&30$\pm5$&5.5\\
&&&&&&&-32.5&2.5&12$\pm4$&- \\
&&&&&&&-45.0&2.5&6$\pm1$&- \\
&&&&&&&-53.5&2.5&5$\pm1$&- \\
\hline
\multicolumn{11}{l}{\sl ** = saturated absorption component} \\
\hline
\end{tabular}
\end{flushleft}
\end{table*}
}

\section{Discussion}

\cite{white85} presented high resolution (but low signal-to-noise) NaI absorption data
towards 28 stars in the direction of Ara~OB1 and have listed three pervasive absorption
components at heliocentric velocities of $-37$, $-18$ and 0\,km\,s$^{-1}$. Tentative
evidence was presented that associated both of the negative velocity components with that
of the Ara~OB1 association, such that the gas clouds were placed within the
Carina-Sagittarius arm at a distance of $\sim 1.3$\,kpc. However,  recent HI
\citep{mcclure02} and H$\alpha$ \citep{georg96} measurements suggest a kinematic distance
$> 2.5$\,kpc for these negative velocity gas components, thus apparently precluding a
physical link with Ara~OB1. Hence, in this present work, we wish better to identify the
distances associated with the major interstellar gas clouds seen in this galactic direction
by using new NaI and CaII absorption data, taken in the context of  data recorded since
1985 at other wavelengths. First we follow the example of \cite{white85} by plotting the
velocity extent of the absorption detected towards each target, with the stars ordered
according to their distance, in Figure 5. This plot reveals four broadly distinct groupings
of absorption components that cover the following velocity ranges towards the following
targets: $-10$\,km\,s$^{-1} < V_{\mathrm{helio}} < +10\,$\,km\,s$^{-1}$ (stars 1 $\&$ 2),
$-20$\,km\,s$^{-1} < V_{\mathrm{helio}} < +20$\,km\,s$^{-1}$ (stars 3, 4, 5 and 11),
$-45$\,km\,s$^{-1} < V_{\mathrm{helio}} < -25$\,km\,s$^{-1}$ (stars 6, 7, 8, 9, 12 and 13)
and $-60$\,km\,s$^{-1} < V_{\mathrm{helio}} < -50$\,km\,s$^{-1}$ (stars 10 and 14). We now
discuss each of these absorption groupings (which we associated with separate interstellar
gas clouds) in order to determine their respective distances and also to comment on the
physical conditions of the ISM present over these four absorption velocity intervals.

\begin{figure*}
\center
\bigskip
\includegraphics[width=12cm]{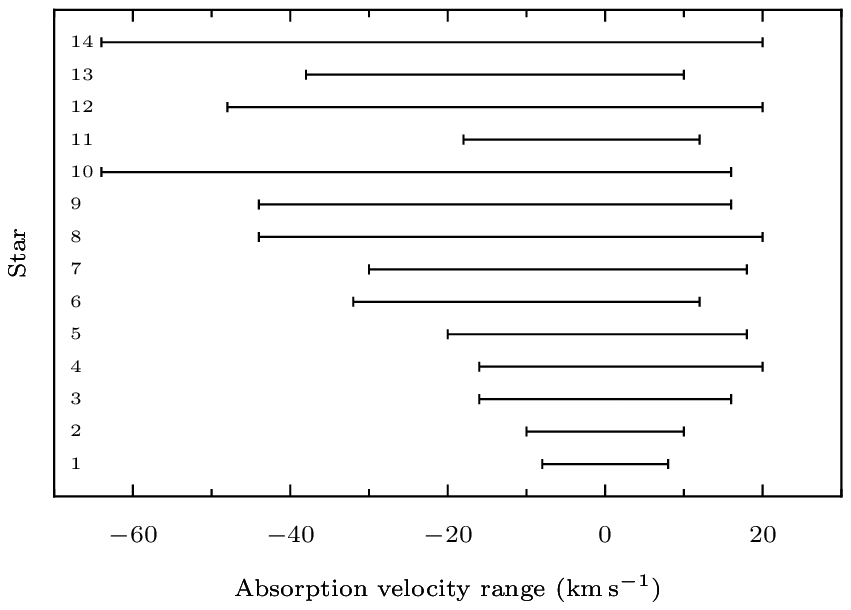}
\caption{The range of absorption velocity in the CaII and/or NaI lines, in order of increasing distance,
measured towards each of the 14 stars (listed by number) in Table 1.}
\label{Figure 5}
\end{figure*}

\subsection{Local interstellar gas ($-10$\,km\,s$^{-1} < V_{\mathrm{helio}} < +10$\,km\,s$^{-1}$)}

Absorption detected along the two sight-lines with distances $< 500$\,pc (HD 147971 and HD
150083) spans the smallest range of velocities in our sample. A similar restricted range of
absorption is shown in the NaI absorption profile of HD 150168 ($d \sim 140$\,pc) by
\cite{white85}. These three sight-lines sample gas primarily associated with the local
interstellar medium, which consists of a very low density region to a distance of $\sim
110$\,pc (i.e.\ the Local Bubble cavity), which is then followed by a cold and dense
neutral gas boundary \citep{lall03}. For most galactic sight-lines the neutral boundary to
the Local Bubble is defined by gas with a NaI D2-line equivalent width of $\sim 50$\,m\AA\
\cite{sfeir99}. Thus, the direction in the galactic plane towards $\it l \sim 337^{\circ}$
is somewhat anomalous in that the neutral boundary gas is about twice as dense as that
found in most other directions. We derive a ratio $N$(NaI)/$N$(CaII) of 5.9 for the
combined gas components detected towards HD 147971, which is a typical value widely found
for gas in the diffuse ($n_{\rm H} \sim 10$\,cm$^{-3}$) ISM \citep{hobbs76}.

Interestingly, the N(NaI) column density towards HD 147971 and HD 150083 differs by only a
factor of $\sim 3.5$ over the $\sim 350$\,pc interstellar path length  between the two
stars, whereas the corresponding $N$(CaII) column density increases by twice this amount
over the same distance. This suggests that there is a significant amount of warm and
ionized gas that fills this interstellar sight-line. This may be explained by the presence
of a very low neutral gas density region of the ISM termed the Lupus-Norma interstellar
cavity \citep{welsh94}, whose approximate dimensions to a distance of 300\,pc have been
mapped by \cite{lall03}.

\subsection{Gas with velocities $-20$\,km\,s$^{-1} < V_{\mathrm{helio}} < +20$\,km\,s$^{-1}$}

The NaI and CaII spectra of the three stars with distances in the 570--800\,pc range (HD
151113, HD~148974 and HD 146444) reveal additional absorption components originating from
gas with velocities in the $-10$ to $-20$\,km\,s$^{-1}$ and  $+10$ to $+20$\,km\,s$^{-1}$
ranges. We note that gas with similar NaI and CaII component velocities of
$V_{\mathrm{helio}} \sim  -10$ to $-20$\,km\,s$^{-1}$ has also been detected in absorption
towards many of the stars of the Sco OB1 association, but only towards those with distances
$\gtrsim 720$\,pc  \citep{crawford89}.  Similarly, \cite{white85} have detected NaI
absorption with a velocity component of $-14$\,km\,s$^{-1}$ towards HD 150041 ($d = 530\pm
200$\,pc). In addition, HI  and H$\alpha$ emission have also been detected at
$V_{\mathrm{helio}} \sim  -16$\,km\,s$^{-1}$ over most of this region by \cite{rizzo94} and
\cite{georg96}. Taken collectively, these data suggest that  a widespread interstellar gas
cloud(s) with a velocity of $\sim -15$\,km\,s$^{-1}$ is present in the direction of Ara~OB1
at a distance of $\gtrsim  530$\,pc. Using the well defined absorption component at
$V_{\mathrm{helio}} \sim$ -13.5\,km\,s$^{-1}$ detected towards HD 146444 as being
representative of this gas, we derive a ratio $N$(NaI)/$N$(CaII) of 0.6. Since the cloud
velocity is quite low we do not attribute such a low ratio value to the effects of grain
destruction by interstellar shocks \citep{routly52}, but instead we favor an overabundance
of (warmer and more ionized) CaII bearing gas in this direction.

We also note the presence of gas with velocities $> +10$\,km\,s$^{-1}$, which is also only
detected for sight-lines with distances $>  550$\,pc. The pervasive nature of this positive
velocity cloud component could mean that its origin may be linked to that of the
$V_{\mathrm{helio}} \sim -15$\,km\,s$^{-1}$ component in the form of it being the receding
wall of a shell of expanding gas. We note that the HI velocity profile of GSH~337+00$-$05
observed by \cite{mcclure02} gives a foreground shell wall velocity of $V_{\mathrm{helio}}
\sim -20$\,km\,s$^{-1}$ and a rear wall velocity of $V_{\mathrm{helio}} \sim
+1$\,km\,s$^{-1}$. Given the general similarity in velocity and distance of the presently
detected gas cloud with that of the GSH~337+00$-$05 shell, we tentatively assign a distance
of $\sim$ 530\,pc for this feature. This would preclude its physical association with the
more distant Ara~OB1 cluster and place it closer to the inter-arm region, or perhaps even
the Lupus-Norma cavity (if it extends to this distance).

\cite{mcclure02} have proposed a simple model for the production of large galactic HI
shells in which the combined effects of the spiral arm density gradient and their inherent
migration velocity could lead to a spatial offset of these shells away from the spiral
shock and into the inter-arm region. Since the galactic inter-arm gas density is very low,
this enables migrating HI shells to expand to large sizes without the need for a local
energy source such as that of stellar winds and/or supernovae. Thus,  the notion that the
large angular size of GSH~337+00$-$05 is due to the expansion of a gas cloud into a low
density (inter-arm) region of the Galaxy rather than it being due to the influence of local
stellar winds and/or a supernova event, would now seem to be supported.

Finally, the value of reddening, $E(B-V) = 0.05$, found for  HD 148974 seems anomalously
low for a star at a distance of $600 \pm 350$\,pc, but the detection of strong CaII
absorption with a velocity $> +10$\,km\,s$^{-1}$ is consistent with its placement at $>
500$\,pc. The apparent lack of interstellar dust in this sight-line (and the apparent
over-abundance of CaII gas) could be due to the presence of the foreground Lupus-Norma
interstellar cavity mentioned previously.

\subsection{Gas with velocities $-45$\,km\,s$^{-1} < V_{\mathrm{helio}} < -25$\,km\,s$^{-1}$}

All of our targets with distances $>  1$\,kpc, except HD 149308, have additional NaI and
CaII absorption features that span the $-25$ to $-45$\,km\,s$^{-1}$ velocity range.
\cite{crawford89} have also detected absorption components with similar velocities towards
all stars with distances $>  1040$\,pc in sight-lines towards the  Sco~OB1 association,
which is close in direction, and \cite{white85}  also detect NaI absorption at these
velocities towards Ara~OB1 stars with distances $\gtrsim  1$\,kpc. \cite{rizzo94} reported
on the detection of a large HI shell towards Ara~OB1 with a central velocity of
$V_{\mathrm{helio}} \sim -35$\,km\,s$^{-1}$, which may be associated with that of H$\alpha$
emission that has been detected with a similar velocity over the whole of this region by
\cite{georg96}. Although \cite{rizzo94} derive a kinematic distance of 2500\,pc for this
gas shell, all of the aforementioned absorption data favour a much closer distance of
$\sim$ 1\,kpc. This would then place this highly negative velocity gas at a distance
similar to that of the Ara~OB1 association, whose central stellar cluster of NGC 6193 has
an average heliocentric radial velocity of $-33.1$\,km\,s$^{-1}$ \citep{arnal88}. The
similarity in these velocities between the gas and the stellar motion strongly suggests
that the cluster stars may presently be embedded within, and traveling through, this
interstellar cloud complex.

Surprisingly, no high negative velocity absorption was detected towards HD 149308 ($d \sim
1380$\,pc). We note that its reddening value of $E(B-V) = 0.31$ is what one might expect
from a star of distance $\sim  900$\,pc. \cite{crawford89} has raised the question that
this star may be under-luminous, and if so then its distance could be as small as 400\,pc.
This would then explain the simple absorption spectrum that we have presently recorded.

\subsection{Gas with velocities $-60$\,km\,s$^{-1} < V_{\mathrm{helio}} < -50$\,km\,s$^{-1}$}

We have detected fast moving gas with a velocity spanning the $-50$ to $-60$\,km\,s$^{-1}$
range towards only two targets, HD 154873 and HD 150958. The velocity of this gas is about
22\,km\,s$^{-1}$ more negative than the radial velocity of the Ara~OB1 cluster stars and
thus this expanding interstellar cloud could be physically linked to the stellar winds
emerging from the central stellar cluster. Recent X-ray observations of the nucleus of NGC
6193 by \cite{skinner05} have revealed the O3-type star, HD 150136, to be one of the most
luminous X-ray sources known, producing a significant shocked stellar wind that strongly
influences the ambient ISM. However, absorbing gas with a velocity in the $-50$ to
$-60$\,km\,s$^{-1}$ range was not detected towards this star, thus suggesting that this
interstellar feature may be located at a distance beyond that of HD 150136 ($d >
1150$\,pc). We also note that NaI and CaII absorption components with a similar velocity
have been observed towards many stars in the more distant (1900\,pc), but close in
direction, Sco~OB1 association by \cite{crawford89}. Similarly, \cite{white85} only
detected gas with this absorption velocity towards HD 151018 and $-47^{\circ}$10941, whose
estimated distances are 3\,kpc and 2.2\,kpc respectively. Given the measure of the
uncertainties in the derived distances towards all of the target stars which have
absorption at this highly negative velocity, we tentatively place a minimum distance to
this gas cloud of 1.5--1.8\,kpc.

A NaI/CaII ratio of 0.2 is found for the absorption component observed at
$V_{\mathrm{helio}}  =  -57$\,km\,s$^{-1}$ towards HD 150958. This ratio value is typical
for interstellar gas clouds with $\mid V \mid > 30$\,km\,s$^{-1}$ and has been interpreted
as interstellar grain material returning to the gas through sputtering caused by
interstellar shocks \citep{siluk74}. The origin of this highly negative velocity
interstellar feature is uncertain, given that its derived distance places it beyond that of
Ara~OB1.  Although the cloud's distance is more compatible with that of the Sco~OB1
cluster, for a projected size of $\sim 6^{\circ}$ on the sky (i.e.\ the distance between HD
154873 and HD 150958) its equivalent diameter of $\sim 180$\,pc would seem too large for a
cloud to be created by stellar winds alone. We believe instead that a more likely
explanation is that we have observed gas with serendipitously similar absorption velocities
that are not physically associated over such a large area of the sky.

\section{Conclusion}

We have made high resolution absorption measurements of the NaI and/or CaII lines towards
14 early-type stars in the galactic direction of the Ara~OB1 cluster, which is thought to
lie at a distance of $\sim 1320$\,pc. The interstellar sight-line to this region is
complex, consisting of many gas clouds with velocities ranging from $-60$ to
$+20$\,km\,s$^{-1}$. Previous radio emission and ground-based absorption measurements were
in disagreement as to the distances to these clouds, particularly that of the large
GSH~337+00$-$05 HI shell.

Our new measurements allow the observed line-profiles to be split into four distinct
groupings of absorption velocity. For gas with absorption covering the velocity range
$-10$\,km\,s$^{-1} < V_{\mathrm{helio}} < +10$\,km\,s$^{-1}$, we can identify the observed
absorption with local gas belonging to the Lupus-Norma interstellar cavity located a
distance of between 100 and 485\,pc. Gas with absorption velocities spanning the range
$-20$\,km\,s$^{-1} < V_{\mathrm{helio}} < +20$\,km\,s$^{-1}$ is only detected towards stars
with distances of 570--800\,pc. We identify the angularly large interstellar absorption
feature at $V_{\mathrm{helio}} \sim -15$\,km\,s$^{-1}$ with the expanding HI shell called
GSH~337+00$-$05, which is now firmly placed at a distance of $\sim  530$\,pc.

Gas with absorption velocities in the range  $-45$ \,km\,s$^{-1} < V_{\mathrm{helio}} <
-25$\,km\,s$^{-1}$ is only detected towards stars with distances $>  1$\,kpc, which places
this gas at a similar distance to that of the Ara~OB1 stellar cluster. The similarity
between the radial velocity of the stellar cluster and that of the absorbing gas suggests
that the cluster stars may be presently embedded within, and traveling through, this
interstellar cloud complex. Absorption at velocities $< -50$\,km\,s$^{-1}$ is seen towards
only two stars, HD 154873 and HD 150958, both of which have distances $>  1300$\,pc.
Although the observed NaI/CaII ratio for this component is consistent with the presence of
interstellar (stellar wind-driven) shocks,  we believe that it is unlikely that this highly
negative velocity gas is physically associated with the Ara~OB1 cluster itself, and its
origin may be linked to the more distant (and larger) Sco~OB1 cluster.

\begin{acknowledgments}

We gratefully acknowledge the staff at the Mt John Observatory for their help in making
these observations possible. BYW acknowledges funding from NSF grant $\#$ AST-0507244.

Figure 1 was produced using data from the Southern Galactic Plane Survey \citep{mcclure05}.

This research has made use of the SIMBAD \\ database, operated at CDS, Strasbourg, France.

This research has made use of the VizieR database \citep{ochsenbein00}, operated at CDS, Strasbourg, France.

\end{acknowledgments}

\end{document}